\documentclass[a4paper]{jpconf}
\usepackage{graphicx}
\usepackage[utf8]{inputenc}
\usepackage{amsmath}
\usepackage{amssymb}
\usepackage[numbers,square,sort&compress]{natbib}
\usepackage{epstopdf}
\begin{document}
\title{The positronium hyperfine structure:\\ Progress towards a direct measurement of the $\text{2}^\text{3}\text{S}_\text{1} \rightarrow \text{2}^\text{1}\text{S}_\text{0}$ transition in vacuum}

\author{Michael W. Heiss, Gunther Wichmann, Andr\'{e} Rubbia, Paolo Crivelli}

\address{Institute for Particle Physics and Astrophysics, ETH Zurich, Otto-Stern-Weg 5, 8093 Zurich, SWITZERLAND}

\ead{michael.heiss@phys.ethz.ch, paolo.crivelli@phys.ethz.ch}

\begin{abstract}
We present the current status for the direct measurement of the positronium hyperfine structure using the $\text{2}^\text{3}\text{S}_\text{1} \rightarrow \text{2}^\text{1}\text{S}_\text{0}$ transition. This experiment, currently being commissioned at the slow positron beam facility at ETH Zurich, will be the first measurement of this transition and the first positronium hyperfine splitting experiment conducted in vacuum altogether. This experiment will be free of systematic effects found in earlier experiments, namely the inhomogeneity in static magnetic fields and the extrapolation from dense gases to vacuum. The achievable precision is expected to be on the order of $10\, \mathrm{ppm}$ while the systematic uncertainty is estimated to be within a few $\mathrm{ppm}$. This would allow to check recent bound state QED calculations and a $3\sigma$ discrepancy with earlier experiments.
\end{abstract}

\section{Introduction}
Positronium is a purely leptonic bound state comprising a positron and an electron. As such it is mostly free from effects stemming from the strong and the weak interaction that protonic atoms exhibit\footnote{Such corrections enter in higher order loop processes, but this is strongly suppressed.} and it is free from finite size effects (i.e. proton charge radius). For this reason and the fact that recoil effects are strongly enhanced in positronium, it represents a unique atomic system to test bound state QED to very high precision \cite{Karshenboim2004}. The hyperfine splitting between the triplet and the singlet state is in particular sensitive to higher order QED corrections, which are known to very good accuracy \cite{Czarnecki1999, Baker2014, Adkins2014, Eides2014, Adkins2014_2, Eides2015, Adkins2015, Adkins2015_2, Adkins2016, Eides2016, Eides2017}. Numerous experiments determining this energy splitting in the ground state have been conducted in the past, the most precise being the measurements by Ritter et al. \cite{Ritter1984} and Mills \cite{Mills1983}. While they were all largely compatible with each other, they show a deviation of more than $3\sigma$ from recent bound state QED calculations.
 
All those experiments share common experimental techniques, which were identified as possible sources of systematic errors \cite{Asai2008}. Due to the significant technical difficulties in producing the HFS transition frequency $\Delta_\text{HFS} \approx 203\, \mathrm{GHz}$, the transition was not measured directly. Instead, it was extracted from the Zeeman splitting measured in a static magnetic field. Therefore, the uncertainty in the uniformity of the magnetic field in the experimental region directly contributes to the systematic error. 

Recently, two experiments were conducted trying to minimize these systematic effects by using a very precisely determined magnetic field \cite{Ishida2012, Ishida2014} or by measuring the transition directly using novel millimeter microwave techniques \cite{Miyazaki2015}. While the former experiment yielded a result favoring the bound state QED calculations over the previous experimental average, a more precise measurement is needed to conclusively check the discrepancy with theoretical results. The latter experiment, while being completely free of this systematic effect, has very limited experimental sensitivity on the permill level due to the technical challenges involved and can therefore only be considered as a proof of principle.

Futhermore, all experiments conducted so far used gas as a target to form positronium. This influences the measured transition frequency $\Delta_\text{HFS}$ due to the Stark effect induced by local electric fields of the gas atoms. To obtain the HFS in vacuum, extrapolation to zero gas density is needed, which in turn introduces another source of systematics.

Therefore, to conclusively resolve the $3\sigma$ discrepancy between previous experiments and QED calculations, a sufficiently precise direct measurement of the transition frequency in vacuum is needed. The status of our experiment, ongoing at ETH Zurich, which aims to measure the hyperfine splitting in positronium to the ppm level is reported here.

\section{Experimental technique} \label{sec:technique}

Since the ground state HFS splitting is approximately $203\, \mathrm{GHz}$, this leads to significant technological challenges. The low $\mathrm{THz}$ range is one of the last remaining largely undeveloped frequency regions, a fact which is referred to as the Terahertz gap \cite{Kleiner2007}. As such, special devices for the production, manipulation and measurement of such radiation have to be individually developed, built and characterized, often resulting in large systematic uncertainties.

However, one can exploit that the transition frequency for different excited states $n$ scales approximately like
\begin{equation}
\Delta_\text{HFS}(n) \approx \frac{\Delta_\text{HFS}(n=1)}{n^3}
\text{ .}
\label{eq:freqscaling}
\end{equation}

\begin{figure}[ht!]
\centering
\includegraphics[trim={5 5 5 5},clip,width=0.6\columnwidth]{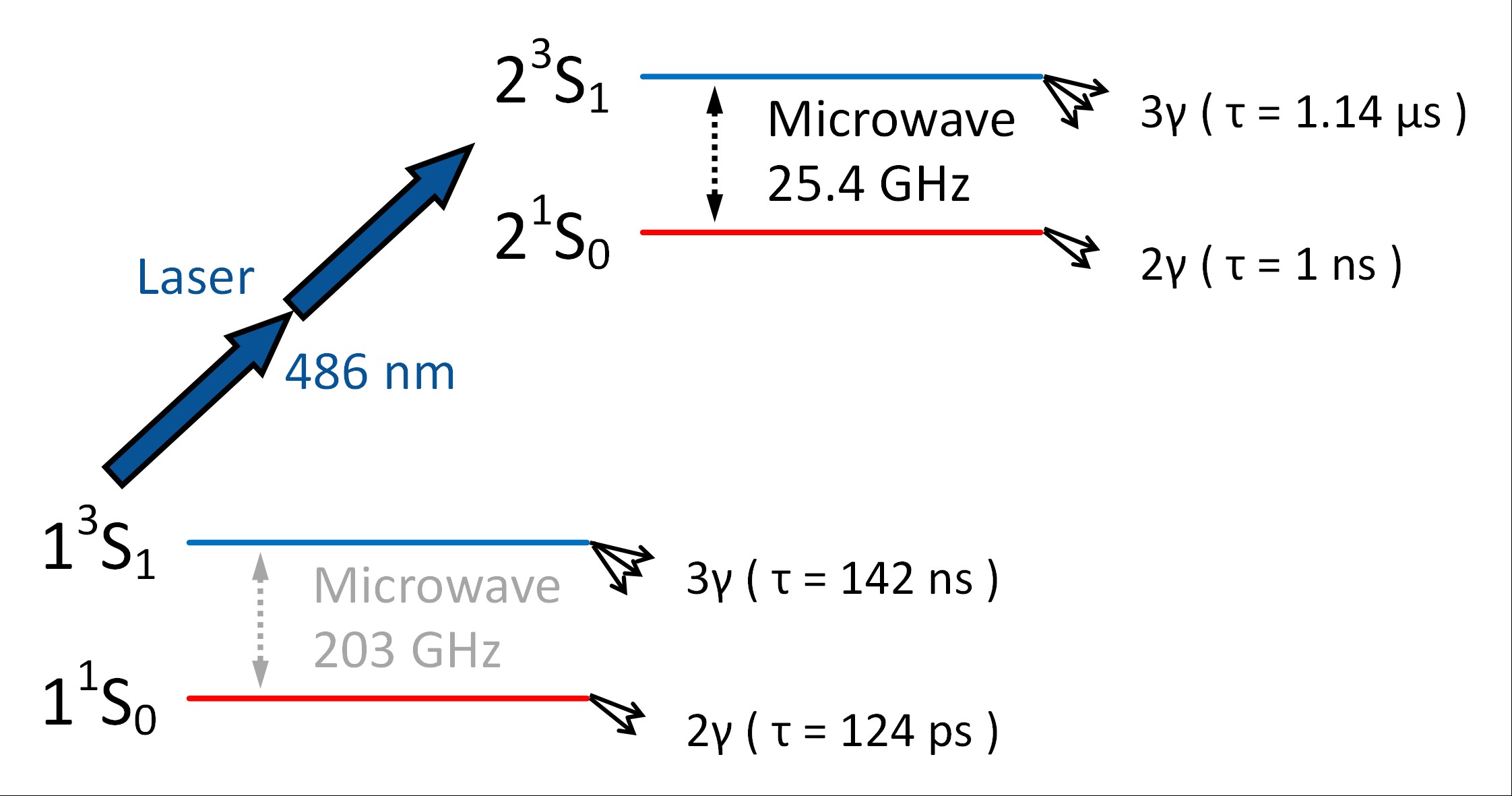}% 
\caption{Overview of energy levels, decay channels and lifetimes: The two-photon laser excitation is shown with bold arrows, while the hyperfine splitting is shown with dotted arrows. Next to the respective hyperfine states the decay channels and lifetimes can be found.}%
\label{fig:level-diagram}%
\end{figure}

If the ground state positronium atoms are excited to the first excited state using lasers, this allows for the $\text{2}^\text{3}\text{S}_\text{1} \rightarrow \text{2}^\text{1}\text{S}_\text{0}$ hyperfine transition to be measured (see figure \ref{fig:level-diagram}). Since the frequency for this transition is around $25.4\, \mathrm{GHz}$, the required microwave field can be generated, manipulated and characterized by using mostly commercially available products.

To efficiently produce positronium in vacuum and achieve sufficient overlap between the laser beam and the produced Ps atoms, one needs to employ a positron beam. Instead of using a gas target as in earlier experiments, a solid target is used, i.e. a porous silica thin film \cite{Crivelli2010,Cassidy2010}. This serves to efficiently convert positrons into positronium by the impinging positron capturing an electron in the bulk of the porous structure, where the penetration depth is dependent on the kinetic energy of the positron, e.g. around 150 nm at 3 keV. The positronium atom then diffuses within the interconnected pore network and is ejected into vacuum.  These positronium atoms can then be laser excited to the $n=2$ state and subsequently the $\text{2}^\text{3}\text{S}_\text{1} \rightarrow \text{2}^\text{1}\text{S}_\text{0}$ hyperfine transition can be induced by using microwave radiation amplified by a resonator.

\begin{figure}[ht!]
\centering
\includegraphics[width=0.65\columnwidth]{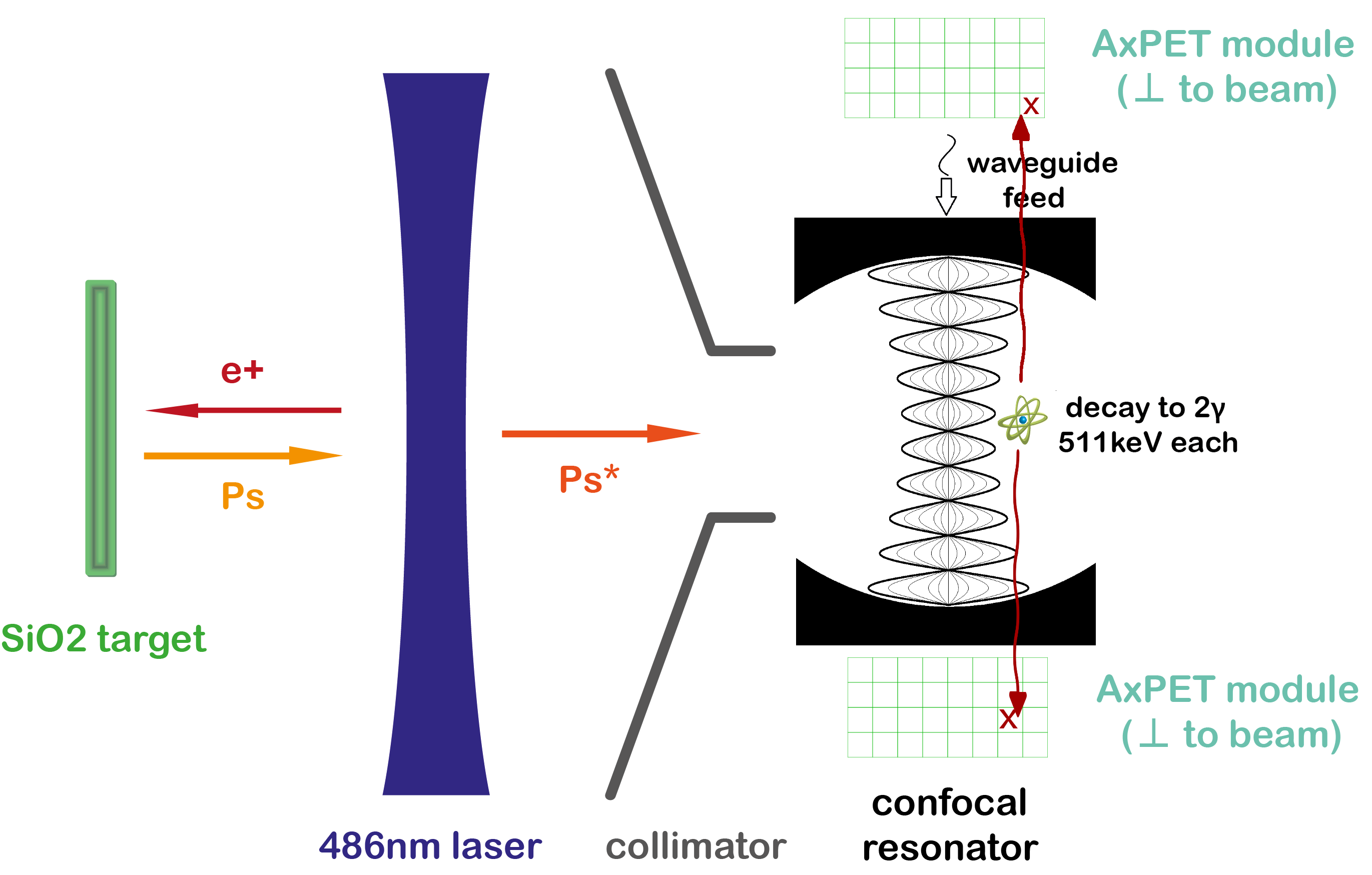}%
\caption{Simplified measurement scheme of the Ps hyperfine transition: Positrons from the bunched positron beam are focused onto the porous silica thin film target. The ejected positronium is then laser excited to the 2S state, spatially filtered and undergoes the hyperfine transition in the microwave field of the resonator. Subsequently, the back-to-back decay photons are picked up by two segmented scintillation detector modules.}%
\label{fig:scheme}%
\end{figure}

Since the singlet $\text{2}^\text{1}\text{S}_\text{0}$ state decays with a lifetime of approximately $1\, \mathrm{ns}$, positronium traverses only around $100\, \mu\mathrm{m}$ and decays within the resonator after the hyperfine transition is induced (see figure \ref{fig:scheme}). The line shape can therefore be measured by distinguishing two-photon decays from three-photon decays with high efficiency and measuring the ratio of the two rates. Finally, fitting the simulated lineshape to the data one can extract $\Delta_\text{HFS}$ and compare it to bound state QED theory results. 

\section{Experimental setup, recent results and current status}

The current effort to precisely measure the $\text{1}^\text{3}\text{S}_\text{1} \rightarrow \text{2}^\text{3}\text{S}_\text{1}$ transition in positronium \cite{Cooke2015, Wichmann2018} at the slow positron beam facility at ETH Zurich offers the opportunity to reuse most of the existing technology for this excited state hyperfine splitting measurement. The pulsed slow positron beam is capable of delivering on the order of $10^5$ positrons per second, coming in $1\,\mathrm{ns}$ bunches at a rate of typically $1\,\mathrm{Hz}$, to a porous silica thin film target \cite{Cooke2016}. The conversion efficiency of positrons to positronium is around $30 \%$ with the atoms emitted according to a $\cos \theta$-distribution. The thermal energies of positronium emitted from porous silica targets are limited by the quantum mechanical ground state energy of the atom in a pore of the target structure. For commonly used pore sizes, this results in a kinetic energy of approximately $40 - 60 \, \mathrm{meV}$ \cite{Crivelli2010,Cassidy2010}.
 
\subsection{Laser excitation}

\begin{figure}[ht!]
\centering
\includegraphics[width=0.8\columnwidth]{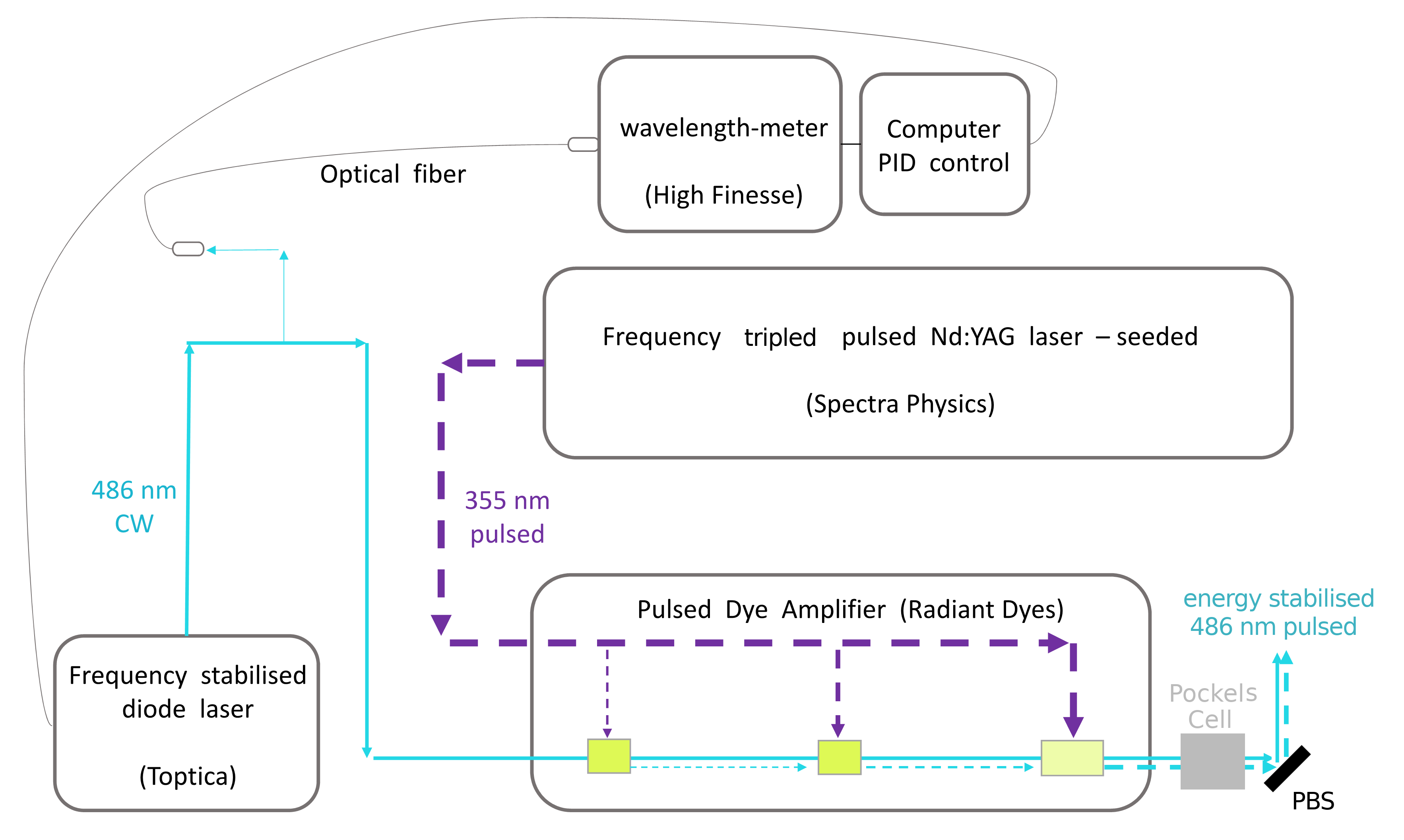}%
\caption{Schematic setup of the laser system: The three used laser wavelengths are produced by an Nd:YAG and two subsequent dye lasers. The dye laser responsible for the 1S-2S transition is further seeded by a Toptica diode laser and stabilized in frequency and energy.}%
\label{fig:laser-setup}% 
\end{figure}

A schematic overview of the laser system can be found in figure \ref{fig:laser-setup}. It consists of a Toptica frequency stabilized CW laser operating at $486 \, \mathrm{nm}$, which corresponds to half the frequency of the transition. The CW laser alone would not result in sufficient intensity to excite efficiently to the $n=2$ state. For this reason an additional Radiant Dyes pulsed dye amplifier is used, which is pumped by a Spectra Physics pulsed Nd:YAG operating in the UV. The laser frequency can be scanned over the required range of a few hundreds of MHz to account for possible shifts by using a PC interface coupled to a wavelength meter. A Pockels cell and a corresponding polarizing beam splitter is placed at the output of the pulsed dye amplifier. This serves both to safely dump the returning laser pulse and to stabilize the long term drift in pulse energy, e.g. due to the degradation of the dye mixture, to below 1\%. This laser setup is capable of delivering pulse energies up to $20\,\mathrm{mJ}$ with pulse lengths of approximately $\text{7}\,\mathrm{ns}$ (full width at half maximum). 

Monte-Carlo simulation predicts that with careful tuning of the laser parameters, the maximum achievable 2S laser excitation probability is on the 1\% level \cite{Heiss2016}. This is mostly limited by direct photo-ionization in the exciting laser and restricted geometrical overlap.

\begin{figure}[ht!]
\centering
\includegraphics[width=0.55\columnwidth,clip,trim=160 260 180 110]{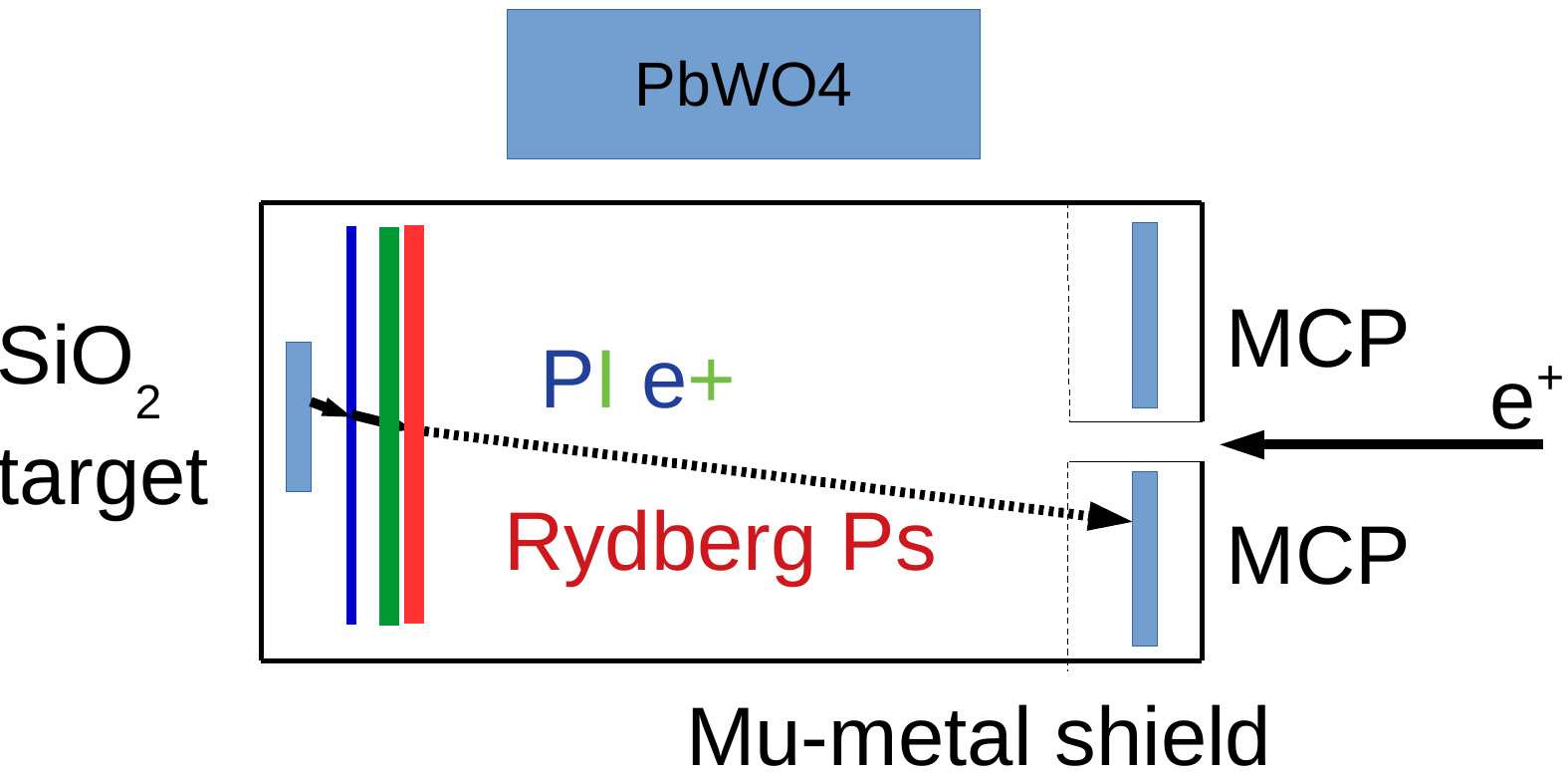}%
\caption{Sketch of positronium excitation and detection scheme: Positrons from the bunched positron beam are focused onto the silica thin film target. The created positronium is then laser excited to the 2S state by the 486nm laser and can subsequently be ioinzed by a 532nm laser or excited to Rydberg states using a 736nm laser. The positrons from photo-ionization or from field-ionization of the Rydberg atoms can either be detected on the MCP or via annihilation photons in a PbWO$_4$ scintillator.}
\label{fig:detection-scheme}% 
\end{figure}

Figure \ref{fig:detection-scheme} shows a schematic overview of the technique to detect the production of positronium and the subsequent excitation to the 2S state. A monolithic $\text{PbWO}_4$ scintillator is placed on top of the experimental chamber to monitor the conversion efficiency of positronium via a technique known as SSPALS \cite{Cassidy2006}. Additionally, the rate of back-scattered positrons is monitored by a micro-channel plate (MCP) placed at approximately $4 \, \mathrm{cm}$ distance, which is directly proportional to the number of incident positrons on the target.

The laser excitation is measured with three independent techniques. When positronium traverses the laser beam, a significant fraction will undergo direct photo-ionization if the frequency is close to resonance. Subsequently, the unbound positrons can either be detected by the MCP or by an excess in the SSPALS spectrum due to the pulsed nature of the exciting laser. However, while this shows that some positronium is excited to the 2S state, it does not constitute a direct measurement of the 2S positronium fraction. This is realized by employing additional pulsed lasers, which are spatially and temporally separated from the 2S exciting laser. 

The first method uses a 532nm Nd:YAG laser pulse with approximately 30mJ in 7ns (FWHM) to ionize the surviving 2S atoms with a probability close to unity. Again, the positron from photo-ionization can be picked up by either the MCP or its annihilation photons in the scintillator.

\begin{figure}[ht!]
\centering
\includegraphics[width=0.75\columnwidth]{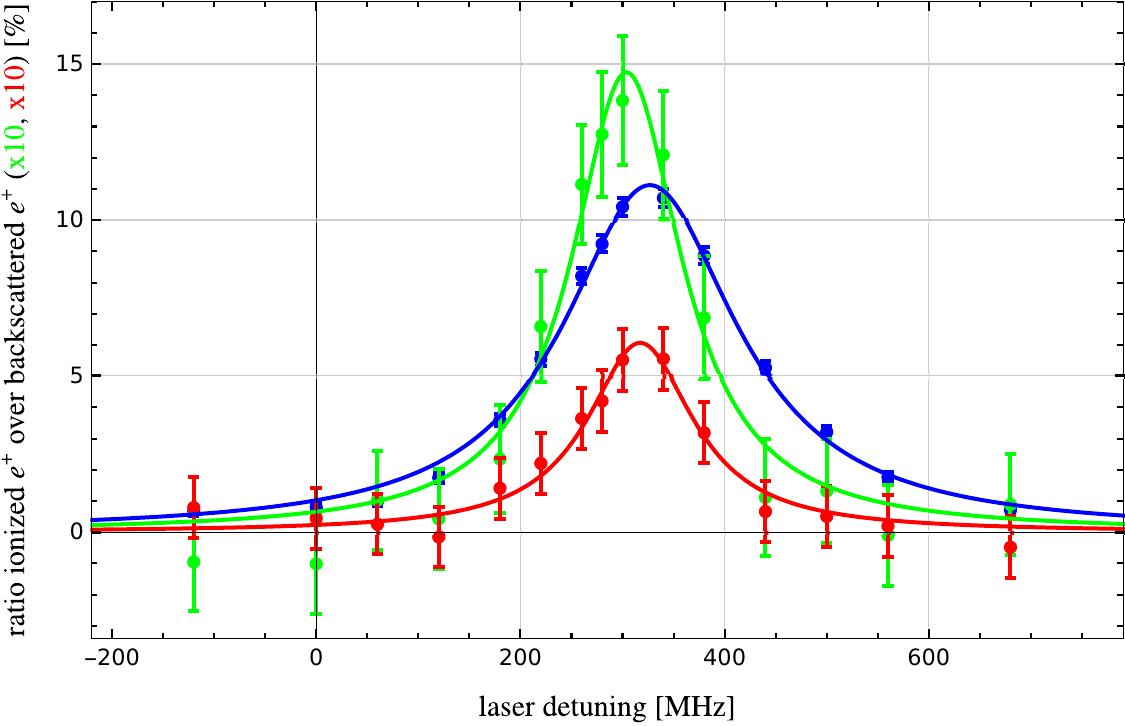}%
\caption{Positronium 2S excitation probability as a function of laser frequency: The measured data is shown as dots with error bars, while the lines are simple Lorentzian fits. Data in blue corresponds to direct photo-ionization in the exciting laser, while green represents photo-ionization in the 532nm laser. The red data points are signal due to field-ionization of Rydberg positronium on the MCP grid.}%
\label{fig:lineshape}% 
\end{figure}

Alternatively, we employ a 736nm dye laser pulse with approximately 2mJ in 7ns (FWHM) to induce the transition from 2S to 20P. When this Rydberg state encounters the high electric field ($\approx 4\,\mathrm{kV/cm}$) between a grounded grid and the front plate of the MCP it is field ionized and the positron is picked up by the MCP. Due to the time of flight of the Rydberg positronium, this signal is delayed by around 400ns and can be used to measure the atoms velocity. This method was developed and will be used for the correction of the second order Doppler shift, the main systematic effect, in the 1S-2S measurement \cite{Cooke2015, Wichmann2018}. Figure \ref{fig:lineshape} shows a typical frequency scan of the 486nm laser exciting positronium to the 2S state, employing these detection techniques. 

\begin{figure}[ht!]
\centering
\includegraphics[width=0.6\columnwidth]{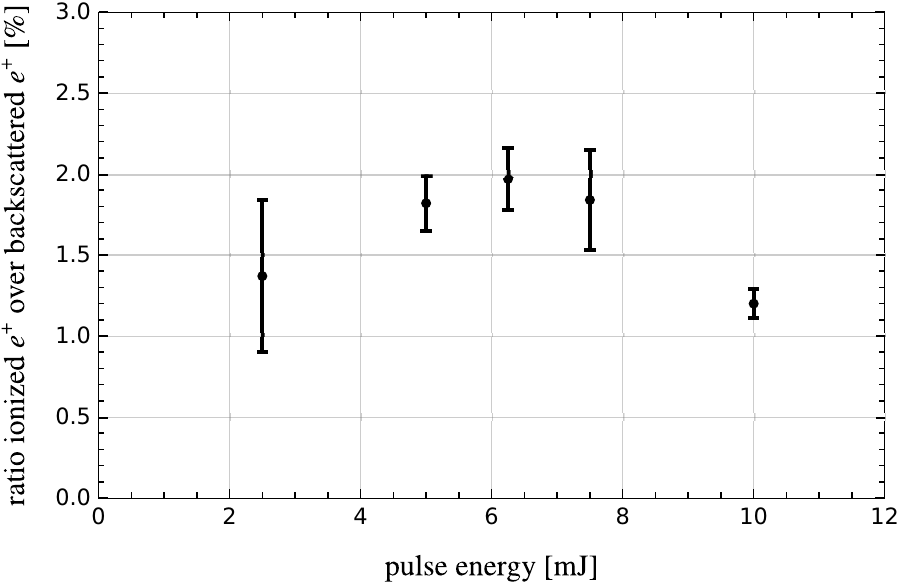}%
\caption{Fraction of 2S excited Ps atoms (normalized to the back-scattered peak)
 as a function of the laser pulse energy.}%
\label{fig:powerscan}% 
\end{figure}

The fraction of surviving positronium atoms in the 2S state was optimized by scanning the available laser parameters, e.g. the temporal and spatial overlap and the pulse energy (see figure \ref{fig:powerscan}). A peak ratio of approximately 2\% of delayed photo-ionized to back-scattered positrons was measured. Multiplying the normalization factor
\begin{equation}
N = \frac{P_\text{BS}}{\varepsilon_c\left(1-P_\text{BS}\right)} \approx 0.26
\text{ ,}
\label{eq:normfactor}
\end{equation}
where $\varepsilon_c \approx 0.25$ is the conversion efficiency of positrons to positronium on the target and $P_\text{BS} \approx 0.06$ is the probability of a positron to back-scatter into the detector predicted by simulation \cite{Vigo2017}, we estimate the fraction of excited and surviving positronium atoms to be
\begin{equation}
P_{2S} \gtrsim 0.5\%
\text{ .}
\label{eq:2sprob}
\end{equation} 
 
\subsection{Microwave transition}

The hyperfine transition is achieved by using a time-varying magnetic field close to the resonance frequency, which is on the order of $25.4\, \mathrm{GHz}$. A custom confocal resonator is used to form a sufficiently strong standing microwave field. The attainable transition rate is limited by the input power $P_\text{in}$, the coupling losses, and the quality factor $Q$ of the cavity. 

\begin{figure}[ht!]
\centering
\includegraphics[width=0.5\columnwidth,clip,trim={100 100 150 140}]{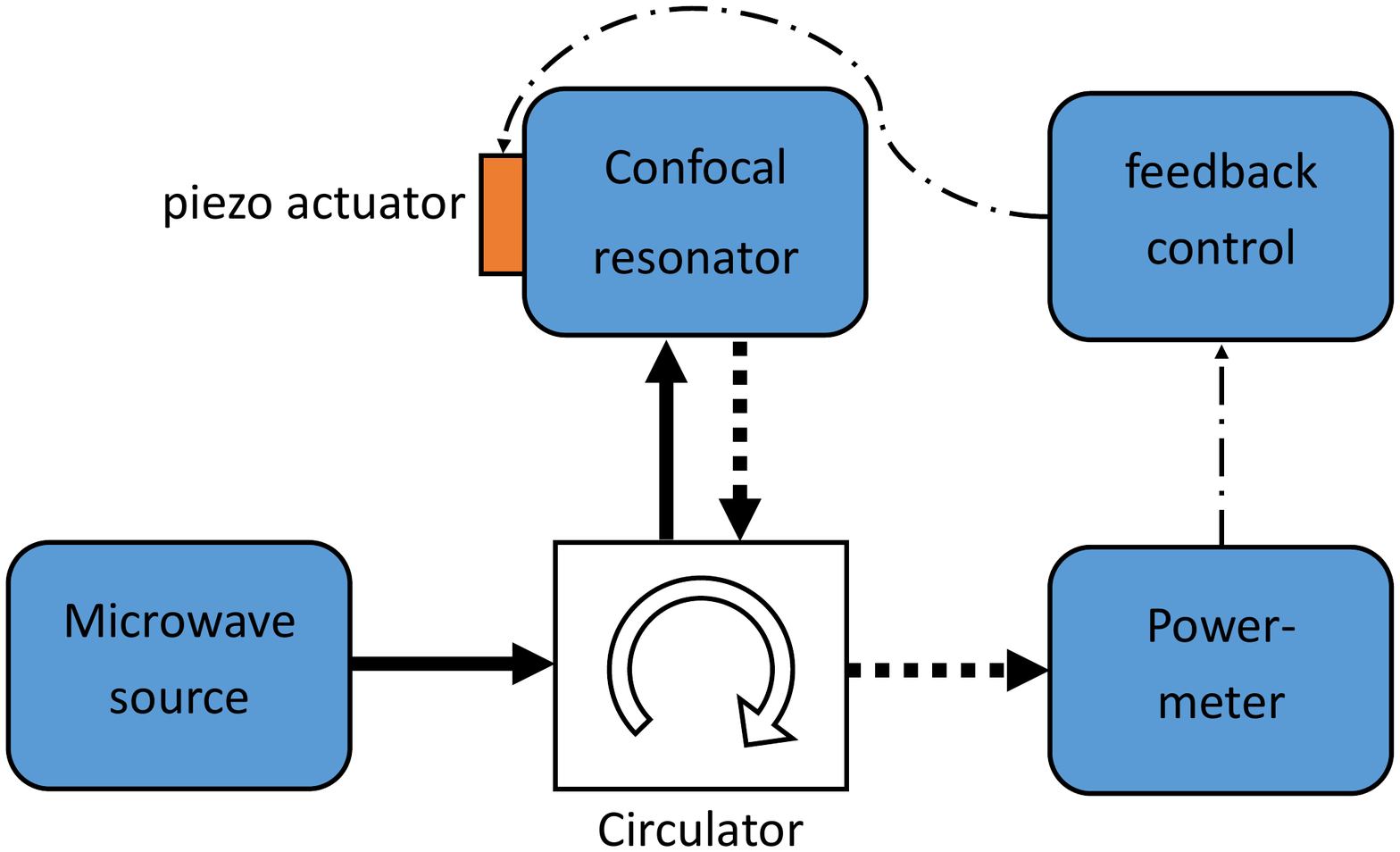}%
\caption{Schematic overview of the cavity monitoring and control system: The microwave radiation is directionally coupled via a circulator into the resonator, while the reflected power is directed into a powermeter. A feedback loop controls the piezo actuator used to stabilize the cavity.}%
\label{fig:cavity-control}% 
\end{figure}

The resonator parameters were chosen such that the only excited mode is the fundamental with design values of a loaded $Q$ factor of approximately 33500 at critical coupling or equivalently a FWHM $\approx 0.8\,\mathrm{MHz}$ of the spectral response \cite{Heiss2016}. This is significantly smaller than the transition line shape, which is broadened due to the short lifetime of positronium to about $160 \, \mathrm{MHz}$ FWHM. Therefore, a piezo actuator is used to lock and stabilize the cavity to the input frequency, by controlling the distance between the two spherical mirrors. The stabilization and monitoring system is based on a simple custom scalar network analyzer (see figure \ref{fig:cavity-control}) with a feedback loop to the piezo actuator, which allows for the cavity to be stabilized to the percent level over several days.

\begin{figure}[ht!]
\centering
\includegraphics[width=0.55\columnwidth]{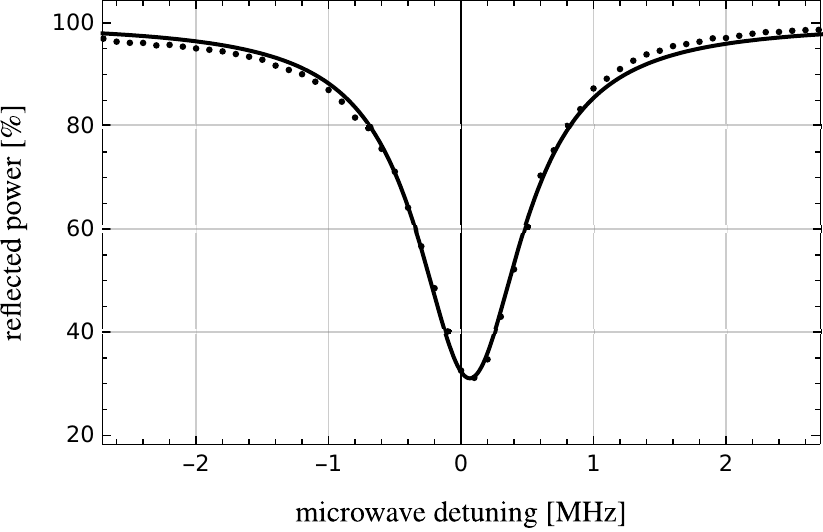}%
\caption{Measured spectral response of confocal resonator including Lorentzian simple fit. The data was calibrated offline to the spectral response of the system without the cavity.}%
\label{fig:cavity-response}% 
\end{figure}

The spectral response of our confocal resonator setup was measured to have a FWHM of about $\Gamma \approx 1\,\mathrm{MHz}$ and $Q\approx 26300$, which are close to the design value. However, as shown in figure \ref{fig:cavity-response}, only about 70\% of the incident power is coupled into the cavity. This is likely due to a small mismatch of the coupling hole size at room temperature (the ideal coupling hole size depends on the unloaded Q and therefore on the resistive losses in the cavity). In the final vacuum setup, the cavity is cooled and stabilized with thermoelectric elements. Furthermore, a second set of mirrors with a smaller coupling hole diameter was produced and is available to be optimized for ideal coupling, e.g. by iteratively widening the coupling hole size via wire erosion.

From simulation we expect a transition probability (including geometrical considerations) of around 0.5\% using the analog signal generator alone ($P_\text{in} \approx 200\,\mathrm{mW}$) and up to approximately 15\% using an additional  microwave amplifier ($P_\text{in} \approx 10\,\mathrm{W}$) \cite{Heiss2016}. 

\subsection{Detection and analysis}

To detect the two back-to-back $511 \, \mathrm{keV}$ photons of the annihilation of the $\text{2}^\text{1}\text{S}_\text{0}$ state and to discriminate against the overwhelming background of three-photon decays of the $\text{2}^\text{3}\text{S}_\text{1}$ and $\text{1}^\text{3}\text{S}_\text{1}$ states, excellent spatial, temporal, and energy resolution is required of a detector. These requirements are very similar for Positron Emission Tomography (PET), a medical imaging technique measuring positron annihilations in biological material. Such a detector has recently been developed, amongst others, by the Group of Prof. Dissertori at ETH Zurich within the AX-PET collaboration as a demonstrator device \cite{Beltrame2011} and is available to be used for this experiment. However, the DAQ needed to be refurbished due to differing requirements and more electrical noise in our experimental environment, which is an ongoing effort.

\begin{figure}[ht!]
\centering
\includegraphics[width=0.40\columnwidth]{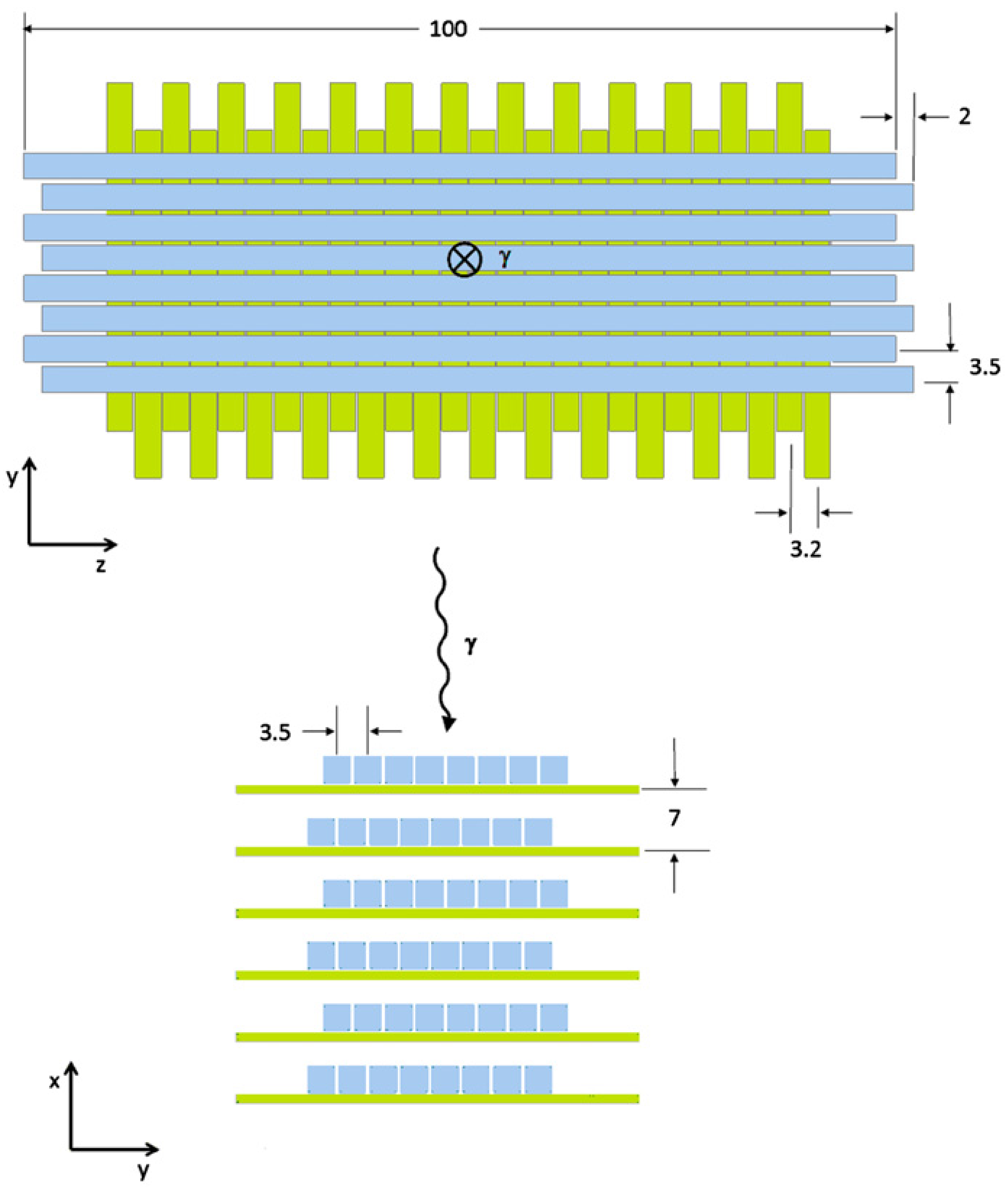}%
\caption[Schematic view of a detector module]{Schematic view of a detector module (\emph{source:} \cite{Beltrame2011}). The detector consists of 6 layers of LYSO crystals interspersed with perpendicular layers of wavelength shifting strips.}%
\label{fig:axpet}%
\end{figure}

Each module consists of two identical modules (see figure \ref{fig:axpet}) built up by six layers of eight $10\, \mathrm{cm}$ long scintillating LYSO crystals each, read out by Geiger-mode Avalanche Photo Diodes. The cross-section of the LYSO crystals of $3 \times 3\, \mathrm{mm}^2$, together with the distance of $3.5 \, \mathrm{mm}$ and $7 \, \mathrm{mm}$ between the crystals, defines the resolution of the detector in the X,Y-plane. Each layer is additionally read out by 26 wavelength shifting strips (WLS) perpendicular to the crystals, connected to G-APDs as well. While the distance between the WLS is $3.2\,\mathrm{mm}$, the precision in the Z-coordinate is on the order of $1\,\mathrm{mm}$, since the signal can usually be seen on multiple WLS and one can use a weighted average to get a better spatial determination \cite{Beltrame2011}.

Background suppression of three-photon decays is implemented via vertex reconstruction and cuts on the deposited energy and timing. Monte-Carlo simulation predicts detection efficiencies for two-photon decays (including solid angle and background suppression considerations) of a few percent, while the signal-to-noise ratio is on the order of 10 \cite{Heiss2016}. Further improvements seem feasible, especially if a neural network analysis approach would be employed.

\section{Summary and conclusion}

In this contribution we illustrated the progress towards the direct measurement of the $\text{2}^\text{3}\text{S}_\text{1} \rightarrow \text{2}^\text{1}\text{S}_\text{0}$ positronium hyperfine transition. The laser excitation and the microwave system were shown to perform close to design specifications. Furthermore, the detector and DAQ were tested and are currently being commissioned.

Simulations and estimations regarding systematic errors predict the sensitivity of the measurement to be on the level of \cite{Heiss2016}:
\begin{equation*}
\Delta\nu_\text{2S-HFS} \,\approx\, \pm \, 10 \, \mathrm{ppm} \text{ (statistical) } \pm \, 4 \, \mathrm{ppm} \text{ (systematic).} 
\label{eq:resultsummary}
\end{equation*}

This would constitute the first measurement of a hyperfine transition of positronium in vacuum, which would be free of systematic effects found in earlier experiments. Furthermore, it could shed some light on a $3\sigma$ discrepancy of bound state QED theory with earlier measurements.

\section*{Acknowledgments}

This work is supported by the SNSF under the grant 16628. P. Crivelli is in debt with F. Merkt for the enlightening discussions on the pulsed laser system, K. Kirch and A. Antoginini for their continuous support, D. Cooke, L. Gerchow, P. Comini and I. El Mais for their essential developments in different stages of this work and L. Laszlo for providing the porous silica thin film targets used for positronium production. Furthermore, the authors like to extend their gratitude to G. Dissertori and W. Lustermann for providing the AxPET demonstrator device and F. Nessi-Tedaldi for lending us the PbWO$_4$ crystals, T. Donner and U. Hollenstein for their help with the laser system and J. Agner, H. Schmutz, B. Zehr and P. Gomez for their technical support.

\bibliography{literature}

\end{document}